\pdfoutput=1
\documentclass[type=preprint,linenumbering=off]{sphenix}

\usepackage{orcidlink}

\usepackage{lipsum}

\begin{document}

\renewcommand{\thefootnote}{$\star$}

\newcommand{\pp}{$p$+$p$\xspace}
\newcommand{\ppbar}{$p$+$\overline{p}$\xspace}
\newcommand{\xj}{\ensuremath{x_\mathrm{J}}\xspace}
\newcommand{\ptn}[1]{\ensuremath{p_{\mathrm{T},#1}}}
\newcommand{\pt}{\ensuremath{p_\mathrm{T}}\xspace}
\newcommand{\dphi}{\ensuremath{\Delta\phi}\xspace}
\newcommand{\pythia}{\textsc{Pythia}\xspace}
\newcommand{\pythiaV}{\textsc{Pythia-8}\xspace}
\newcommand{\herwig}{\textsc{Herwig}\xspace}
\newcommand{\herwigV}{\textsc{Herwig-7.3}\xspace}
\newcommand{\geant}{\textsc{Geant~4}\xspace}
\newcommand{\fastjet}{\textsc{FastJet}\xspace}
\newcommand{\sqrts}{\ensuremath{\sqrt{s}}\xspace}

\newcommand{\meanpt}{\ensuremath{\langle p_\mathrm{T} \rangle}\xspace}
\newcommand{\sigpsi}{\ensuremath{\sigma(p_{\mathrm{T},\psi})}\xspace}
\newcommand{\siglam}{\ensuremath{\sigma(p_{\mathrm{T},\lambda})}\xspace}
\newcommand{\ptpsi}{\ensuremath{p_{\mathrm{T},\psi}}\xspace}
\newcommand{\ptlam}{\ensuremath{p_{\mathrm{T},\lambda}}\xspace}

\title{Measurement of dijet transverse momentum imbalance and azimuthal acoplanarity in $p$+$p$ collisions at $\sqrt{s} = 200$~GeV with the sPHENIX detector}
\author{sPHENIX Collaboration\footnote{See the appendix for the list of collaboration members}}
\date{\today}

\doctag{sPH-JET-2026-01}
\docdoi{-}


\maketitle

\begin{abstract}
This Letter reports on measurements of dijet transverse momentum (\pt) imbalance and azimuthal acoplanarity in proton--proton collisions at $\sqrt{s} = 200$~GeV, using data recorded by the sPHENIX detector at the Relativistic Heavy Ion Collider corresponding to an integrated luminosity of $41$~pb$^{-1}$. Jets are reconstructed using the anti-$k_t$ algorithm with radius parameters $R = 0.3$ to $0.8$ from electromagnetic and hadronic calorimeter energy deposits.
The jet \pt resolution is determined directly in data using two independent methods.
The dijet \pt imbalance is characterized by the ratio $\xj = \ptn{2}/\ptn{1}$ where $p_\mathrm{T,1(2)}$ is the highest (second-highest) jet \pt in the event. 
The dijet azimuthal acoplanarity $\dphi = |\phi_1 - \phi_2|$ is also reported.  
Results are reported for different $\ptn{1}$ selections and jet radius parameters, normalized per dijet pair, and compared to the results of \pythia and \herwig Monte Carlo event generators. 
These measurements provide a stringent quantitative test of the modeling of QCD parton shower and hadronization dynamics, place important constraints on event-generator descriptions at RHIC energies, and establish a comprehensive proton--proton baseline for forthcoming measurements of jet modification in heavy ion collisions.
\end{abstract}

At hadron colliders such as the Relativistic Heavy Ion Collider (RHIC) and the Large Hadron Collider (LHC), jets are the dominant high transverse momentum (\pt) final state. They are produced largely from $2\to2$ parton scattering processes which, after the subsequent quantum chromodynamics (QCD) evolution, emerge as jet pairs approximately back-to-back in azimuth, referred to here as ``dijets''. In proton--proton (\pp) collisions, measurements of dijet production rates and kinematic correlations serve as tests of perturbative QCD calculations~\cite{Gao:2023ulg}, are used for the extraction of parton distribution functions~\cite{Ethier:2020way}, and provide information for the tuning of Monte Carlo (MC) event generators~\cite{Buckley:2009bj}. Fully corrected dijet kinematic correlations have been measured extensively in \pp collisions at the LHC~\cite{ATLAS:2024png,CMS:2024hwr} and \ppbar collisions at other facilities~\cite{UA1:1983zip,D0:2009rxw}, but not yet at RHIC.

In collisions of heavy nuclei, which create a large and long-lived region of Quark-Gluon Plasma (QGP), dijet production rates and kinematic correlations become modified as the evolving parton showers interact with the expanding QGP medium in a process known as jet quenching~\cite{Wang:2025lct}. 
Measurements of dijet production have historically played an important role in the study of event-by-event jet quenching at RHIC and the LHC~\cite{ATLAS:2010isq,CMS:2011iwn,STAR:2016dfv}, providing input on the path-length dependence of jet quenching (i.e. where the two jets lose different amounts of energy)~\cite{Qin:2010mn,Young:2011qx}, the role of jet-by-jet fluctuations~\cite{Milhano:2015mng}, and the kinematic redistribution of quenched energy~\cite{CMS:2015hkr}. For such measurements, high-quality \pp data provide the essential vacuum reference, capturing baseline decorrelation effects from the parton shower, hard initial- and final-state radiation (ISR and FSR), and other higher-order processes. Only then can the additional physics of the QGP modification be quantitatively interpreted.

The sPHENIX experiment at RHIC~\cite{Belmont:2023fau}, designed for measurements of high-\pt probes of the QGP, is instrumented with electromagnetic (EMCal) and hadronic (HCal) calorimeter systems for measuring jet production over a broad kinematic range. The large pseudorapidity ($|\eta|<1.1$) and full azimuthal acceptance of the calorimeters results in a large fraction of dijet pairs above $\pt > 30$~GeV contained in the detector.    Due to the smaller relative contribution from higher-order QCD processes at RHIC, dijets have a tighter kinematic correlation in \pp collisions than at the LHC, increasing the sensitivity of these observables to a variety of QGP medium effects in central Au+Au collisions~\cite{PhysRevD.99.034006}. Previous measurements of dijet production at RHIC by the STAR experiment~\cite{STAR:2016dfv} were performed using charged-particle tracks and EMCal energy deposits only, included a high-tower trigger bias, and were not unfolded for detector effects.

This Letter reports measurements of dijet \pt imbalance and azimuthal correlations in \pp collisions at $\sqrts = 200$~GeV with the sPHENIX detector, using a dataset corresponding to an integrated luminosity of $41$~pb$^{-1}$. The results are characterized by the \pt ratio $\xj = \ptn{2}/\ptn{1}$, where $p_\mathrm{T,1(2)}$ correspond to the highest (second highest) jet \pt, and by the azimuthal acoplanarity $\dphi = |\phi_1 - \phi_2|$. The \xj distributions are corrected for detector effects via a two-dimensional Bayesian unfolding procedure, while the \dphi distributions are corrected using simulation-derived bin-by-bin correction factors. Results are reported for anti-$k_t$~\cite{Cacciari:2008gp} jet radius parameters from $R = 0.3$ to $R = 0.8$ in steps of $0.1$ and in different ranges of $\ptn{1}$, normalized per dijet pair, and compared to the predictions from \pythia~\cite{Bierlich:2022pfr,Sjostrand:2014zea} and \herwig~\cite{Bewick:2023herwig73,Bellm:2015jjp} event generators.


A detailed description of the sPHENIX detector is available in Refs.~\cite{Sphenix:TDR,sPHENIX:2017lqb,Aidala:2020toz}.
The measurements presented here use the Minimum Bias Detector (MBD), the EMCal and HCal systems, and the Level-1 trigger system.
The MBD provides the minimum-bias trigger signal and the reconstructed collision vertex position, $z_\mathrm{vtx}$. It is located at forward rapidity, $3.51 < |\eta| < 4.61$, on both sides of the interaction point and comprises $64$ quartz radiator tubes on each side. The calorimeter system comprises three radially concentric cylindrical layers with full azimuthal coverage and $|\eta| < 1.1$.  The EMCal is a tungsten-scintillating fiber sampling calorimeter with tower granularity $\Delta\eta \times \Delta\phi = 0.025 \times 0.025$, approximately $20$ radiation lengths deep.   The Inner and Outer HCals Calorimeters (IHCal and OHCal) are aluminum-scintillator and steel-scintillator sampling calorimeters, respectively, with tower granularity $0.1 \times 0.1$. The full calorimeter system totals nearly five hadronic interaction lengths. The EMCal energy response is calibrated tower-by-tower using the $\pi^0$ and $\eta \to \gamma\gamma$ invariant mass, while the HCal response is calibrated using cosmic-ray muons and temperature-dependent corrections~\cite{sPHENIX:2025nbg}.

Events were selected for recording using the Level-1 hardware trigger. Jet triggers were defined using the sum of EMCal and HCal energies within overlapping, sliding windows of approximately $\Delta\eta \times \Delta\phi = 0.8 \times 0.8$, with multiple thresholds. The analyzed data were recorded by requiring a coincidence of the jet trigger with a $10$~GeV threshold and the minimum-bias trigger, which required at least one MBD tube above threshold on each side.


The dataset was recorded during RHIC Run-$24$ in June--October $2024$ and comprises two running periods: a primary period with zero beam crossing angle and a wide $z_\mathrm{vtx}$ distribution ($\sigma_z\sim60$~cm), and a secondary period with a $1.5$~mrad crossing angle and narrower distribution ($\sigma_z\sim20$~cm). Events were required to have a vertex within $|z_{\textrm{vtx}}| < 60$~cm and a collision time consistent with the beam crossing.

Simulations from two MC event generators were used to determine corrections for detector effects and provide generator-level comparisons. \pythia~$8.312$~\cite{Bierlich:2022pfr} was used with the Detroit tune~\cite{Aguilar:2021sfa}, which is optimized for jet observables at RHIC energies. Multiple thresholds on the minimum scattering \pt were combined to populate the physically-falling jet spectrum. Separately, \herwig~$7.3$~\cite{Bewick:2023tfi} was used with the Nashville tune~\cite{Qureshi:2024eqz}. Particle-level jets are defined by applying the anti-$k_t$~\cite{Cacciari:2008gp,Cacciari:2011ma} algorithm to all final-state particles, excluding muons and neutrinos. Simulated events were propagated through the full sPHENIX detector simulation using \geant~\cite{GEANT4:2002zbu}, including emulation of calorimeter noise at levels determined from data, and then reconstructed and calibrated identically to data. 


Jets are reconstructed from calibrated calorimeter towers. The inputs to the jet finder are the energies of individual $\Delta\eta \times \Delta\phi = 0.1 \times 0.1$ IHCal and OHCal towers and EMCal pseudo-towers combined into regions matching the HCal segmentation, which are treated as massless pseudojets. 
The jet energy scale (JES) is corrected by first using a simulation-derived calibration factor as a function of \pt and $R$, determined by matching reconstructed jets to the particle-level jets in \pythia simulation. An additional correction, used to align the response in data and simulation, is then determined \textit{in situ} utilizing $\gamma$+jet and multi-jet \pt balance, following the general procedure used by the ATLAS experiment~\cite{ATLAS:2017bje}.

The jet energy resolution (JER) is determined directly in data using two independent methods~\cite{SupplementalMaterial}. In the bisector method~\cite{UA2:1984tmm,ATLAS:2012cse}, the dijet vector sum $\vec{p}_\mathrm{T} = \vec{p}_{\mathrm{T},1} + \vec{p}_{\mathrm{T},2}$ is decomposed along axes that bisect and are perpendicular to the dijet opening angle. Since the perpendicular component is much more sensitive to detector resolution while both components are similarly sensitive to initial-state radiation, the quadrature difference of their widths isolates the detector contribution. In the dijet imbalance method~\cite{ATLAS:2012cse}, the width of the \pt asymmetry, $A_\mathrm{J} = (\ptn{1} - \ptn{2})/(\ptn{1} + \ptn{2})$, is measured with a veto on additional jets and extrapolated to zero third-jet $\pt$, with the particle-level width subtracted in quadrature. Both methods give consistent results, indicating that the JER ranges from $23$\% at $p_\mathrm{T} \approx 20$~GeV to $16\%$ at $p_\mathrm{T} \approx 50$~GeV. The JER is found to be underestimated in simulation by an amount which can be modeled as an additional uncorrelated 10\% resolution effect, which is applied to the simulation samples.


In data, events are selected if the two highest-\pt jets satisfy $\ptn{1} > 18.3$~GeV and $\ptn{2} > 8.2$~GeV (at the calibrated scale), both lie within $|\eta| < 1.1 - R$ to ensure full containment, and are approximately back-to-back in azimuth, $\dphi > 3\pi/4$. The \xj distribution is obtained via a projection of a corrected two-dimensional $(\ptn{1}, \ptn{2})$ distribution, following the procedure developed in Ref.~\cite{ATLAS:2017xfa}. 
An unfolding procedure utilizing the iterative Bayesian algorithm~\cite{DAgostini:2010hil} in the \textsc{RooUnfold} package~\cite{Adye:2011gm} corrects for \pt bin migration, migration into and out of the fiducial $\eta$ and $\dphi$ selections, and the trigger efficiency. The simulated \pythia samples, including the data-driven JER smearing, are used to populate the response matrix associating particle-level $(\ptn{1}, \ptn{2})$ values with reconstructed ones. The response matrix is reweighted at the particle-jet \pt level using results from an initial unfolding to reduce differences between the $(\ptn{1}, \ptn{2})$ distributions in data and simulation. 

After the reweighting, the full unfolding procedure is performed, with the number of iterations chosen to minimize the quadrature sum of the statistical uncertainties and the iteration-to-iteration changes. The resulting two-dimensional distribution is projected to one-dimensional \xj distributions in three $\ptn{1}$ ranges: $23.9$--$31.2$~GeV, $31.2$--$40.7$~GeV, and $40.7$--$60.8$~GeV. 
The procedure is validated using a half-sample closure test in simulation, with the unfolded and original particle-level distributions agreeing within the statistical uncertainties of the samples.

\begin{figure*}[!th]
    \centering
    \includegraphics[width=0.95\linewidth]{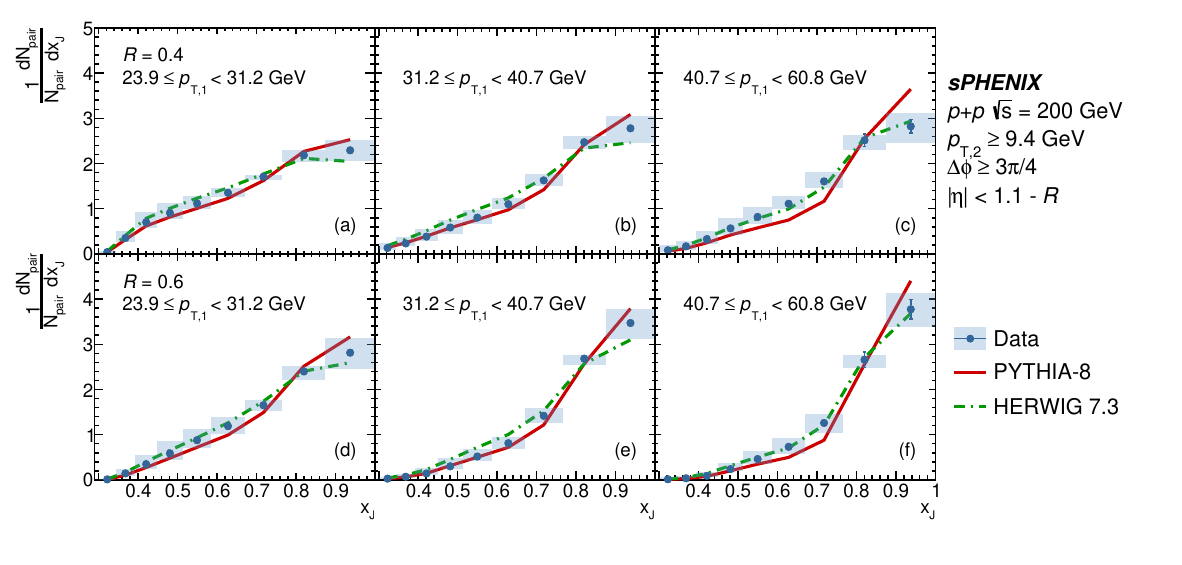}
    \caption{Unfolded \xj distributions for $R = 0.4$ (upper row -- panels (a)-(c)) and $R=0.6$ (lower row -- panels (d)-(f)) jets in three $\ptn{1}$ selections. Vertical bars indicate statistical uncertainties, and boxes indicate systematic uncertainties. Data are compared to particle-level \pythia (Detroit tune) and \herwig (Nashville tune) calculations. 
    }
    \label{fig:xj_result}
\end{figure*}

As the bin migration effects for \dphi\ are small, the \dphi distributions are corrected using bin-by-bin multiplicative factors, derived from the ratio of particle-level to reconstructed-level \dphi distributions in \pythia events.   




The main sources of systematic uncertainty in the measurement are the JES, JER, and the unfolding procedure. 
The JES uncertainty of $\pm 2$--$5$\% is determined from the \textit{in situ} calibration, with a smaller uncertainty at lower \pt{} where the $\gamma$+jet sample has high statistics. The response in simulation is modified according to this range and the unfolding repeated. A significant portion of the JES uncertainty cancels in \xj since it is a ratio of two jet \pt values.
The JER uncertainty is evaluated by considering the agreement between the bisector and dijet imbalance methods, the sensitivity to analysis choices, and the statistical precision of the data-simulation comparisons. The response in simulation is alternatively smeared with a relative JER differing from the nominal by $\pm2$\%, determined from the comparison between the two methods and the statistical uncertainties in the data, and the unfolding is repeated. The JER uncertainty is the dominant uncertainty for the \xj measurement.
The sensitivity of the unfolding to the physics model is assessed by comparing results obtained using \pythia and \herwig for the response.
An uncertainty for potential defects in the unfolding procedure is determined from the half-sample closure test described above. 
For the \dphi measurement, the uncertainty was taken to be half the size of the full correction applied to data, which is co-dominant along with the JES and JER.


Figure~\ref{fig:xj_result} shows the unfolded \xj distributions for $R = 0.4$ and $R = 0.6$ jets in three $\ptn{1}$ selections, compared to particle-level \pythia and \herwig calculations. The distributions are peaked at $\xj = 1$, with the peak rising and narrowing for higher $\ptn{1}$ selections, as expected from the reduced phase space for large-angle radiation at higher \pt. The data are reasonably described  by both \pythia with the Detroit tune and \herwig with the Nashville tune, with \pythia (\herwig) slightly over (under) predicting the yields at large \xj.

\begin{figure*}[!th]
    \centering
    \includegraphics[width=0.95\linewidth]{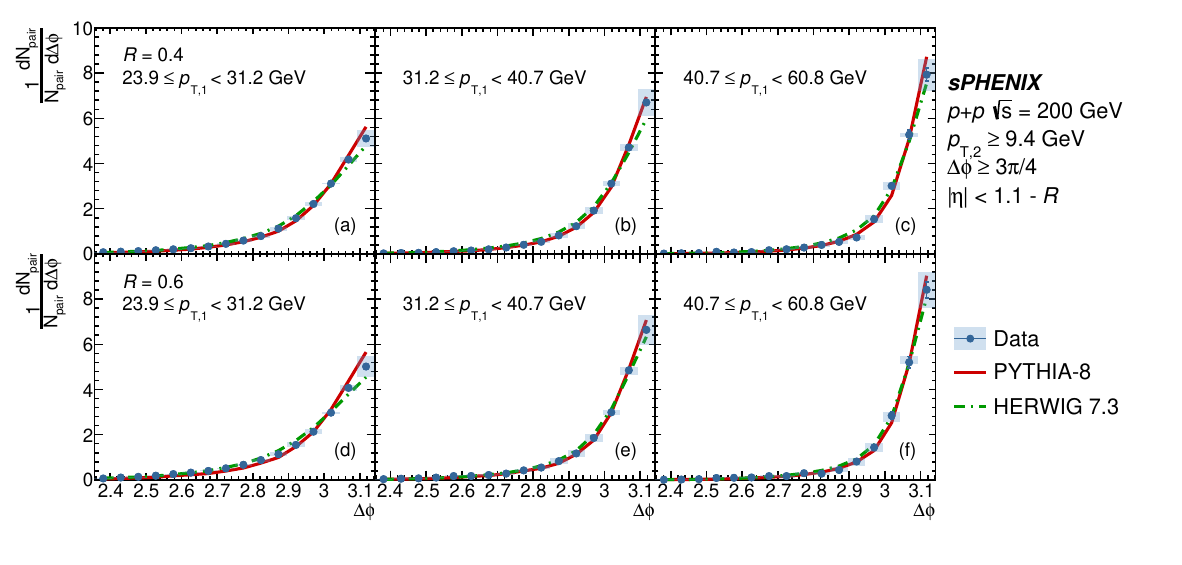}
    \caption{Corrected \dphi distributions for $R = 0.4$ (upper row -- panels (a)-(c)) and $R=0.6$ (lower row -- panels (d)-(f)) jets in three $\ptn{1}$ selections. Vertical bars indicate statistical uncertainties, and boxes indicate systematic uncertainties. 
    Data are compared to particle-level \pythia (Detroit tune) and \herwig (Nashville tune) calculations.
    }
    
\label{fig:dphi_result}
\end{figure*}

Figure~\ref{fig:dphi_result} shows the corrected \dphi distributions for $R = 0.4$ and $R = 0.6$ jets. The distributions are sharply peaked at $\dphi = \pi$, reflecting the dominant back-to-back topology, with a sharper peak at higher $\ptn{1}$. Both \pythia and \herwig qualitatively capture the angular correlation between the dijet pairs in data.

\begin{figure}[t]
    \centering
    \includegraphics[width=0.45\linewidth]{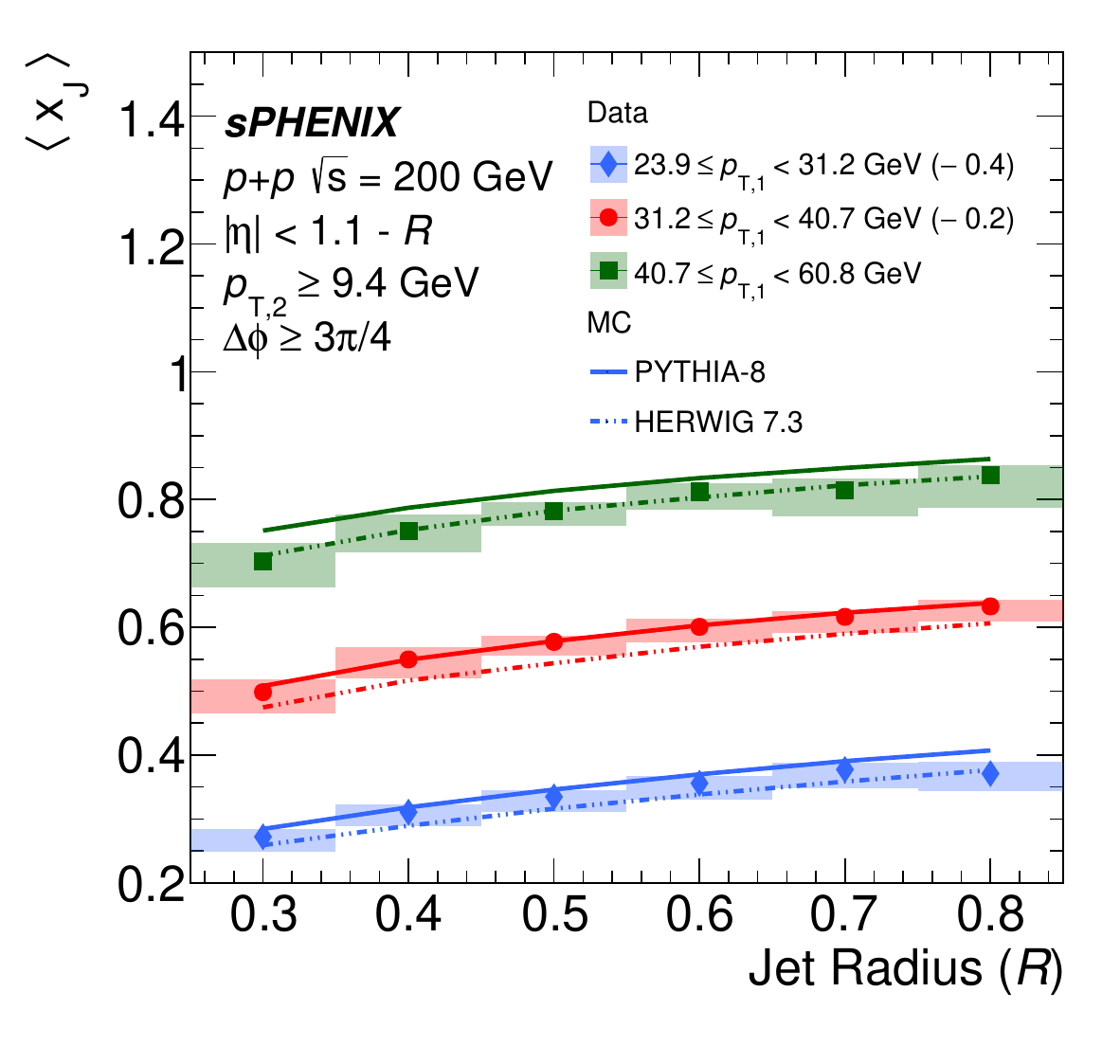}
    \includegraphics[width=0.45\linewidth]{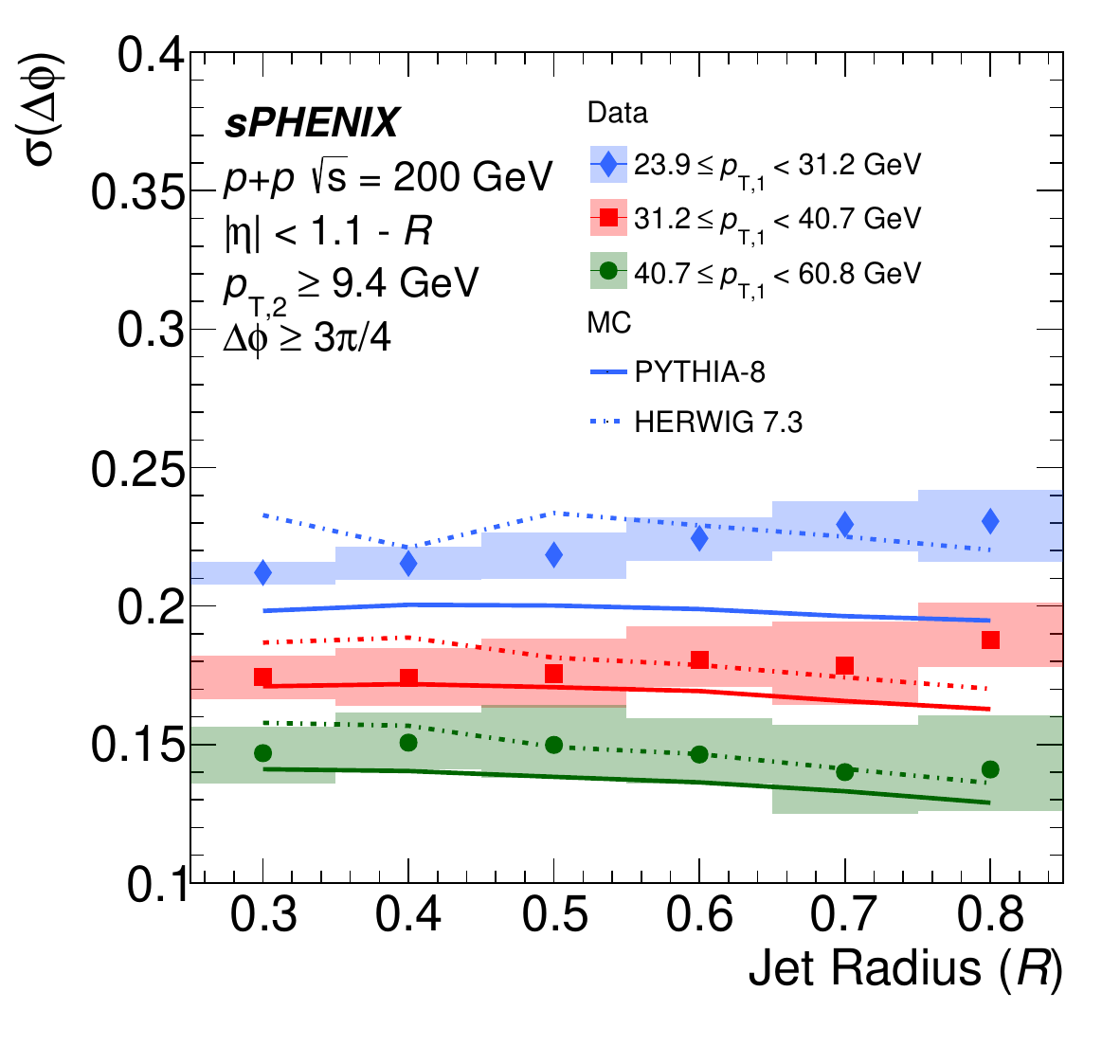}
    \caption{Mean \xj, $\langle \xj \rangle$ (top) and $\sigma(\dphi)$ (bottom) as a function of jet radius $R$ for the three $\ptn{1}$ selections. In the top panel, the different $\ptn{1}$ selections are shifted for clarity.  Vertical bars indicate statistical uncertainties, and boxes indicate systematic uncertainties.  Data are compared to \pythia (Detroit tune) and \herwig (Nashville tune) calculations.}
    \label{fig:meanxj}
\end{figure}

Figure~\ref{fig:meanxj} gives a quantitative summary of the jet-size dependence of the results by showing the mean $\xj$, $\langle \xj \rangle$, and the root-mean-square of the \dphi distributions around $\pi$, $\sigma(\dphi)$, as a function of the jet radius parameter $R$ for all three $\ptn{1}$ selections.  The $\langle \xj \rangle$ increases with higher \ptn{1} selection and with larger jet $R$ value.    These trends reflect the more balanced \pt{} values in higher-\pt{} dijets where the jet area captures a larger fraction of the total parton shower. In \pythia simulations, no significant $R$-dependence is introduced by the changing $\eta$ acceptance for each $R$ value. The \pythia{} and \herwig{} values are in good agreement with these trends.   For the same reason, the \dphi distribution narrows for higher $\ptn{1}$ selections.     However, the data are consistent with a slight rise or no dependence of $\sigma(\Delta\phi)$ for larger jet radii.    In contrast, both generators have a modestly decreasing jet radius dependence.  Notably, \pythia{} significantly under-predicts the $\sigma(\dphi)$ for the lowest \ptn{1} selection.

\begin{figure}[h!]
    \centering
    \includegraphics[width=0.5\linewidth]{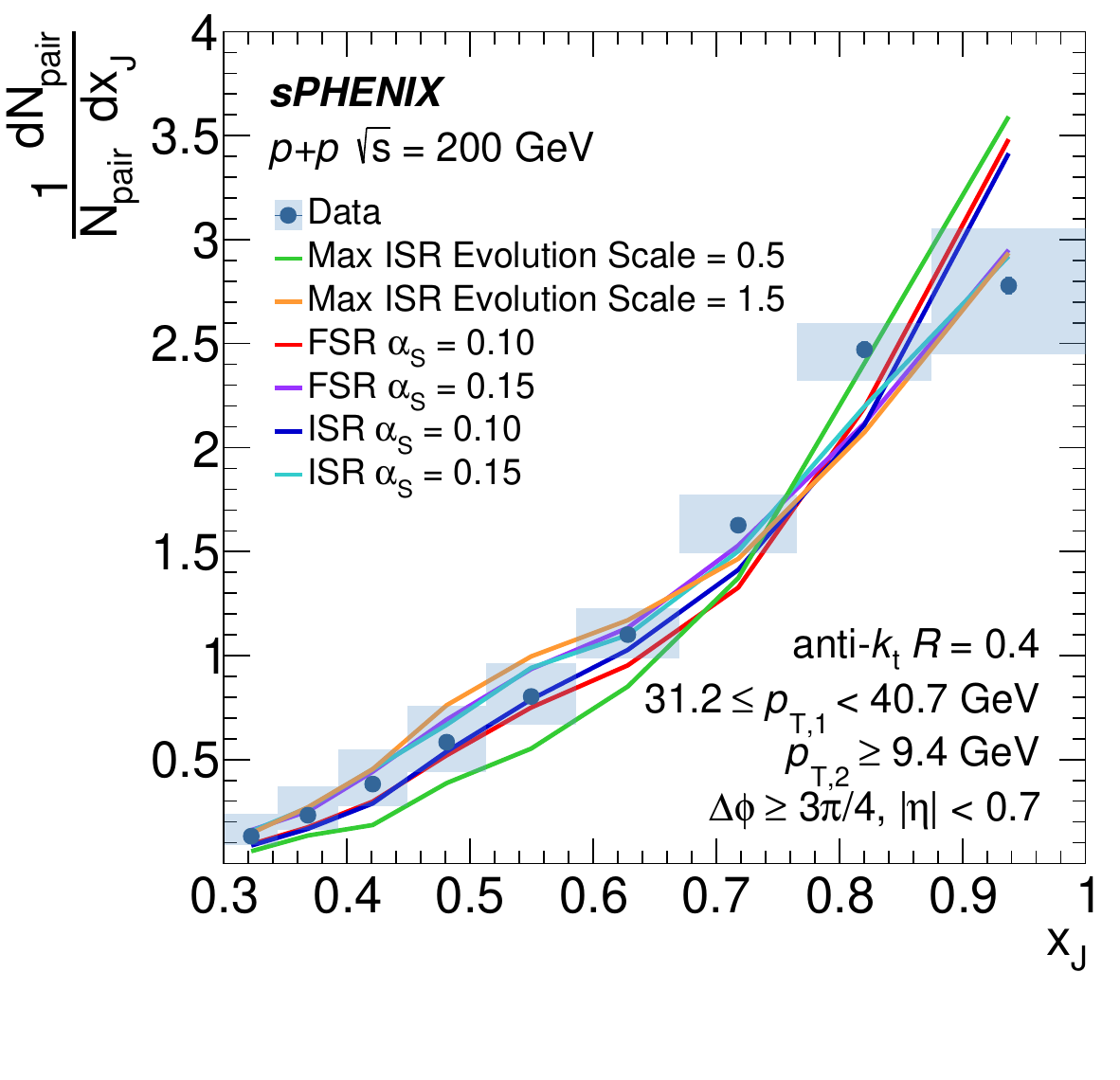}
    \caption{\xj{} distribution for $R=0.4$ jets in the $31.2 \leq \ptn{1} < 40.7$~GeV selection. The data points are shown with systematic uncertainties, along with the distributions from \pythia-8{} configured with minimum/maximum values of parameters sensitive to the description of ISR and FSR modeling at the LHC, holding other parameters fixed.
    }
    \label{fig:pythiaherwigcompare}
\end{figure}

Tuning MC event generators is an important practice at the LHC~\cite{TheATLAScollaboration:2014rfk,CMS:2022awf}, where accurate models of the physics are critical for, e.g., beyond Standard Model searches~\cite{Flesher:2020kuy}. The existing \pythia Detroit tune~\cite{Aguilar:2021sfa} is based on the older Monash tune~\cite{Skands:2014pea} and focuses on the description of the underlying event. To evaluate the discriminating power of the data for descriptions of di-jet kinematic correlations, \pythia{} was run with different parameters which control aspects of the ISR and FSR modeling. Fig.~\ref{fig:pythiaherwigcompare} shows \xj{} distributions, under an example kinematic and jet size selection, for the minimum/maximum values of parameters considered in tuning to jet data at the LHC in Ref.~\cite{TheATLAScollaboration:2014rfk}. The uncertainties on the sPHENIX data are comparable to the envelope set by the different parameter variations, establishing the use of the multi-differential data presented here for precise generator tuning at RHIC.

In summary, this Letter presents measurements of dijet transverse momentum imbalance ratio (\xj) and azimuthal acoplanarity (\dphi) in \pp collisions at $\sqrts = 200$~GeV using data from the sPHENIX detector at RHIC. Jets are reconstructed using the electromagnetic and hadronic calorimeter system with radius parameters $R = 0.3$ to $0.8$. The jet \pt energy scale and resolution are calibrated directly in data using \textit{in situ} $\gamma$+jet and multijet methods and two independent dijet-based methods, respectively. 
The fully corrected \xj and \dphi distributions are qualitatively described by the \pythia and \herwig generators. The \pt and cone-size dependence provides a quantitative test of the modeling of QCD parton shower and hadronization dynamics at RHIC energies, including new constraints for detailed tuning of modern event generators.
These \pp results establish the essential baseline for upcoming sPHENIX measurements of dijet correlations in Au+Au collisions, where modifications to the \xj and \dphi distributions from parton--medium interactions in the QGP are expected. This well-controlled baseline will enable precise studies of the interplay between jet quenching and the recovery of radiated energy at different angular scales, providing new constraints on energy-loss mechanisms in the QGP at RHIC energies~\cite{DEramo:2018eoy,Mehtar-Tani:2021fud,Belmont:2023fau}.



\section*{Acknowledgements}

We thank the Collider-Accelerator Division, Scientific Computing and Data Facilities (SCDF) and Physics Departments at Brookhaven National Laboratory and the staff of the other sPHENIX participating institutions for their vital contributions. We acknowledge support from the Office of Nuclear Physics and Graduate Student Research (SCGSR) program in the Office of Science of the U.S. Department  of Energy, the U.S. National Science Foundation, the National Science and Technology Council, the Ministry of Education of Taiwan, and the Ministry of Economic Affairs (Taiwan), the Ministry of Education, Culture, Sports, Science, and Technology and the  Japan Society for the Promotion of Science (Japan), Basic Science Research Programs through NRF funded by the Ministry of Education and the Ministry of Science and ICT (Korea) and the Swedish Research Council, VR (Sweden).



\bibliographystyle{unsrturl}
\bibliography{references}

\appendix   

\clearpage


\section*{End matter}

\subsection*{Determination of the jet energy resolution in data}


Below we provide a detailed description of the data-driven determination of the jet transverse momentum (\pt) resolution used in the dijet measurement described in the accompanying Letter. The jet \pt resolution, or jet energy resolution (JER), is the dominant systematic uncertainty in the measurement of the dijet momentum balance, \xj, and must be accurately characterized. Two independent methods are employed to determine the JER directly in data and are compared to the resolution in \pythia~\cite{Bierlich:2022pfr,Sjostrand:2014zea} Monte Carlo (MC) simulation, processed through the full \geant~\cite{GEANT4:2002zbu} detector simulation and reconstruction chain. Both methods use dijet events where the two highest-\pt jets have $\pt > 10$~GeV, lie within the pseudorapidity acceptance $|\eta| < 1.1 - R$, and are approximately back-to-back in azimuth, $\dphi > 3\pi/4$.

\subsection*{Bisector method}

The first method, referred to as the bisector method~\cite{UA2:1984tmm,ATLAS:2012cse}, exploits the geometry of the dijet system in the transverse plane. The dijet \pt vector sum, $\vec{p}_\mathrm{T} = \vec{p}_{\mathrm{T},1} + \vec{p}_{\mathrm{T},2}$, is decomposed into two orthogonal components: \ptlam, directed along the axis that bisects the opening angle between the two jets, and \ptpsi, directed along the axis perpendicular to it. Since the $\psi$ axis is approximately collinear with each jet, while the $\lambda$ axis is approximately perpendicular to them, the component \ptpsi measures the difference of two large projections of the individual jet momenta and is thus sensitive to the detector resolution on the jet \pt. In contrast, \ptlam measures the sum of two small components and is much less sensitive to the resolution. However, physics effects that produce a dijet imbalance, such as initial-state radiation (ISR), which is approximately isotropic, are expected to contribute similarly to the widths of both the \ptpsi and \ptlam distributions. 
Thus, the quadrature difference of the two widths isolates the contribution from the detector resolution. 

For each bin in average dijet \pt, the \ptpsi and \ptlam distributions are fitted with Gaussian functions to extract their widths. 
Since the presence of additional jets in the event from hard ISR or final-state radiation (FSR) is explicitly accounted for in this method, there is no additional selection targeting these topologies. However, to ensure that additional jets do not arise from the splitting of a single parton shower into multiple close-by jets, events with a $\pt > 5$~GeV jet within $\Delta R < 1.0$ of either of the two highest-\pt jets are rejected.

The relative jet \pt resolution is then extracted as:
\begin{equation}
\frac{\sigma(p_\mathrm{T})}{p_\mathrm{T}} = \frac{\sqrt{\sigpsi^2 - \siglam^2}}{\sqrt{2}\,\meanpt\,\sqrt{\langle|\cos\dphi|\rangle}}\,,
\label{eq:bisector}
\end{equation}
where $\meanpt$ is the average \pt of the two jets and the $\cos\dphi$ factor accounts for the exact geometry of the dijet system. The potential limitation of this method is that ISR/FSR effects may contribute unequally to the \ptpsi width. This has been studied using \pythia simulations and found to be small. The $\dphi$ resolution of reconstructed dijets is less than the size of a single calorimeter tower across the full \pt range, confirming that the JER dominates the quadrature difference.

\begin{figure}[t]
    \centering
    \includegraphics[width=0.5\linewidth]{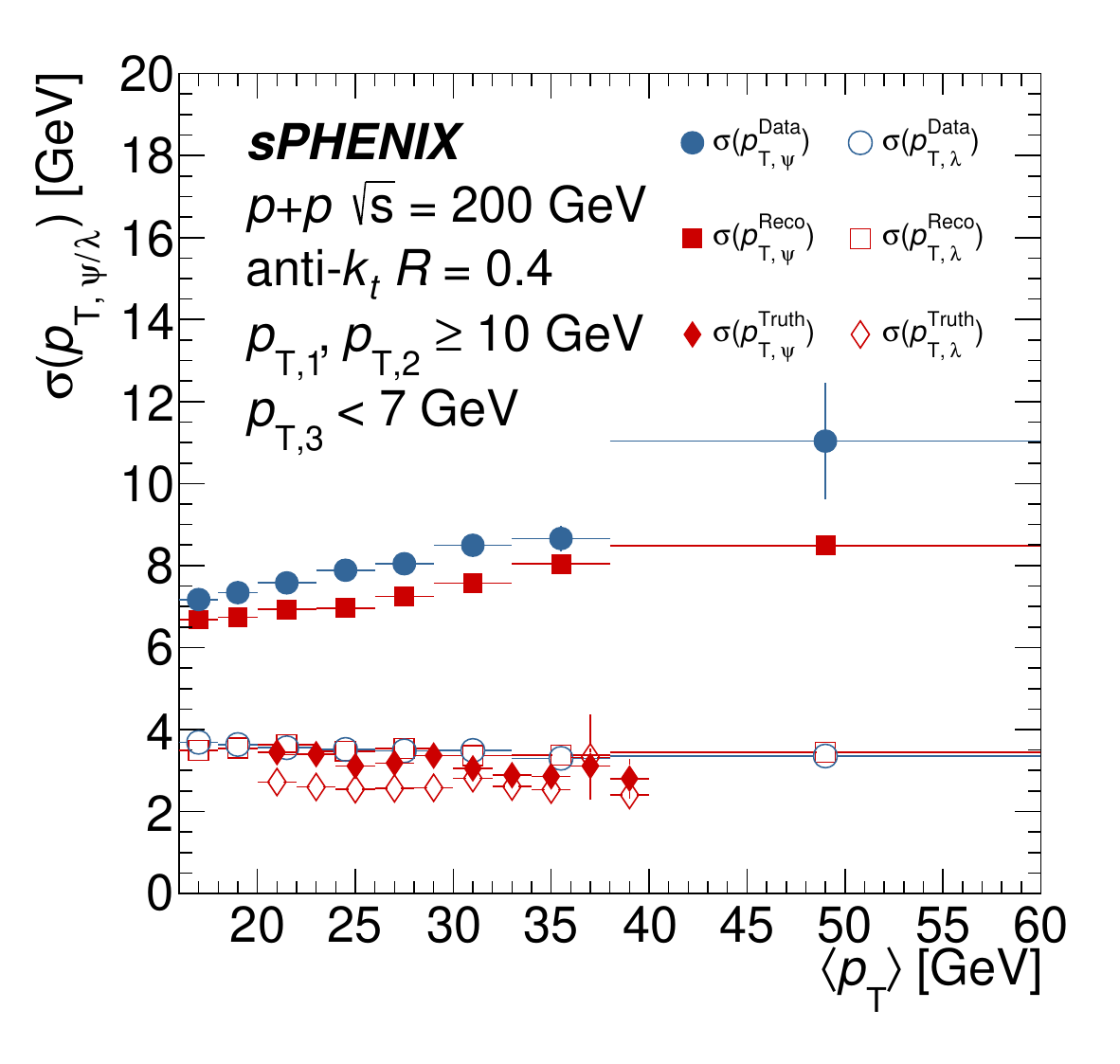}
    \caption{Widths $\sigpsi$ (filled markers) and $\siglam$ (open markers) from the bisector method as a function of $\meanpt$ for $R = 0.4$ jets, shown for generator-level \pythia (diamonds), reconstructed-level \pythia (squares), and data (circles).}
    \label{fig:bisector_summary}
\end{figure}

Figure~\ref{fig:bisector_summary} shows the extracted widths $\sigpsi$ and $\siglam$ as a function of $\meanpt$ for generator-level \pythia, reconstructed-level \pythia, and reconstructed-level data, for $R = 0.4$ jets. At the generator level, the two widths are similar, validating the assumption that physics effects contribute approximately equally along both axes. At the reconstructed level in simulation, the widths increase, with $\sigpsi$ growing faster due to its stronger sensitivity to detector resolution effects. In data, the widths are systematically larger still, indicating an additional contribution to the jet \pt resolution beyond what is described in the simulation.

\subsection*{Dijet imbalance method}

The second method, the dijet imbalance method~\cite{ATLAS:2012cse}, uses the \pt asymmetry, $A_\mathrm{J} = (\ptn{1} - \ptn{2})/(\ptn{1} + \ptn{2})$. In the limit where no additional jets are present in the final state, the width of the $A_\mathrm{J}$ distribution arises from two sources: the detector resolution and the intrinsic imbalance at the particle level from partial containment of the  parton shower in the finite jet cone. To suppress the contribution from hard radiation producing additional jets, the $A_\mathrm{J}$ distribution is constructed with a veto on any third jet above a given $\ptn{3}$ threshold. Since this veto cannot be applied with an arbitrarily low threshold, the $A_\mathrm{J}$ width is evaluated for several values of $\ptn{3}$ and linearly extrapolated to the limit $\ptn{3} \to 0$.

The relative jet \pt resolution is estimated as $\sqrt{2}$ times the quadrature difference between the width of the reconstructed-level $A_\mathrm{J}$ distribution (in data or simulation) and that of the generator-level $A_\mathrm{J}$ distribution in \pythia. The subtraction of the width at particle level accounts for the intrinsic physics contribution to the asymmetry, leaving only the detector resolution contribution. This method relies on the \pythia modeling of the particle-level $A_\mathrm{J}$ width, in contrast to the bisector method which is largely self-calibrating. Thus the complementary modeling assumptions of the two methods provide a valuable cross-check.

\begin{figure}[t]
    \centering
    \includegraphics[width=0.5\linewidth]{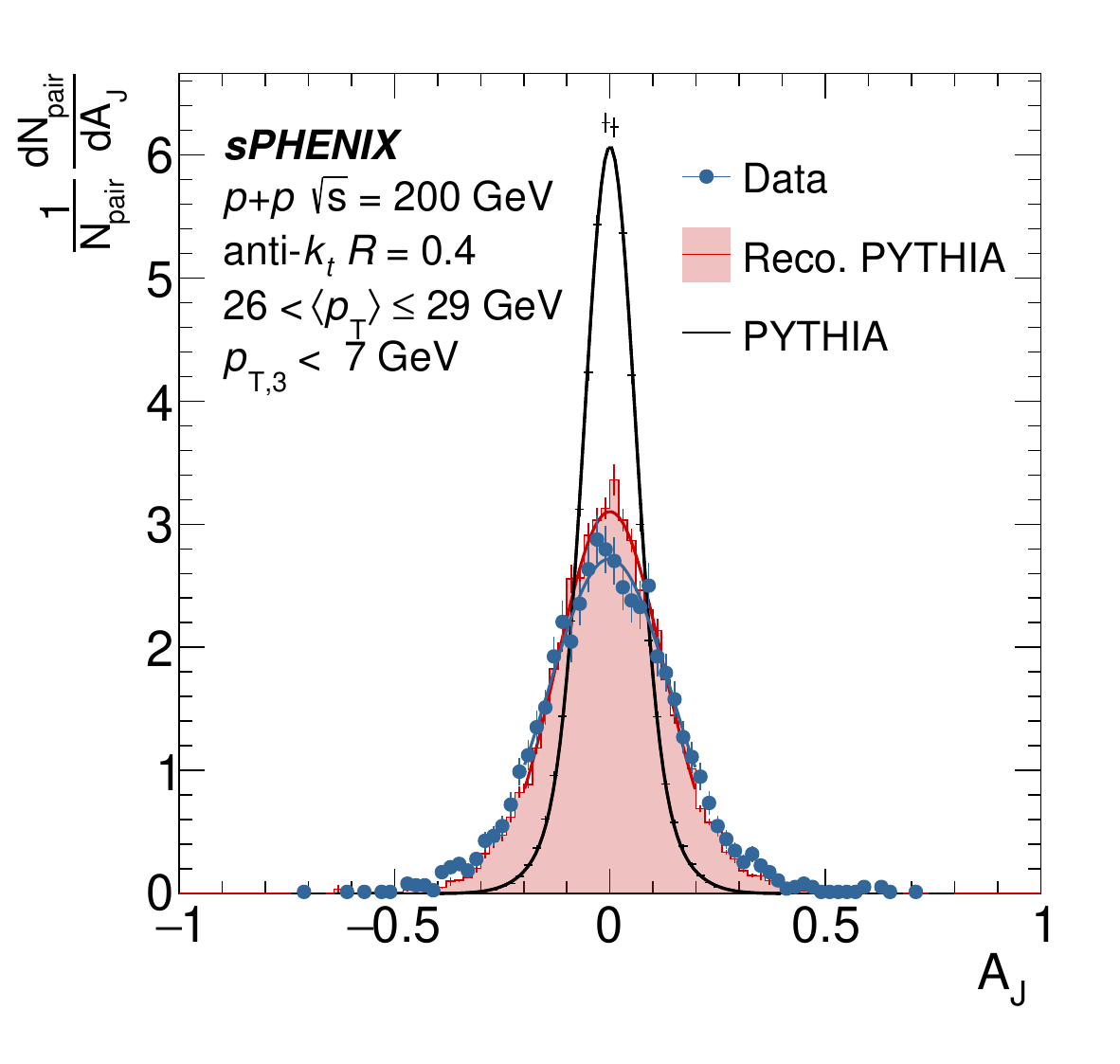}
    \caption{Dijet asymmetry, $A_\mathrm{J}$, distributions for $R = 0.4$ jets in an example $\meanpt$ bin with a third-jet veto threshold of 7~GeV, shown for generator-level \pythia (black), reconstructed-level \pythia (red), and data (blue). The curves show Gaussian fits to the distributions.}
    \label{fig:aj_example}
\end{figure}

Figure~\ref{fig:aj_example} shows the $A_\mathrm{J}$ distributions in an example $\meanpt$ bin for a third-jet veto threshold of $\ptn{3} = 7$~GeV. The distribution has a finite width at the particle level in \pythia, which is then broadened at the reconstructed level by detector resolution. The distribution in data is broader still, consistent with an additional resolution component not captured in the simulation.

\subsection*{Combined results and application to the measurement}

\begin{figure}[t]
    \centering
    \includegraphics[width=0.5\linewidth]{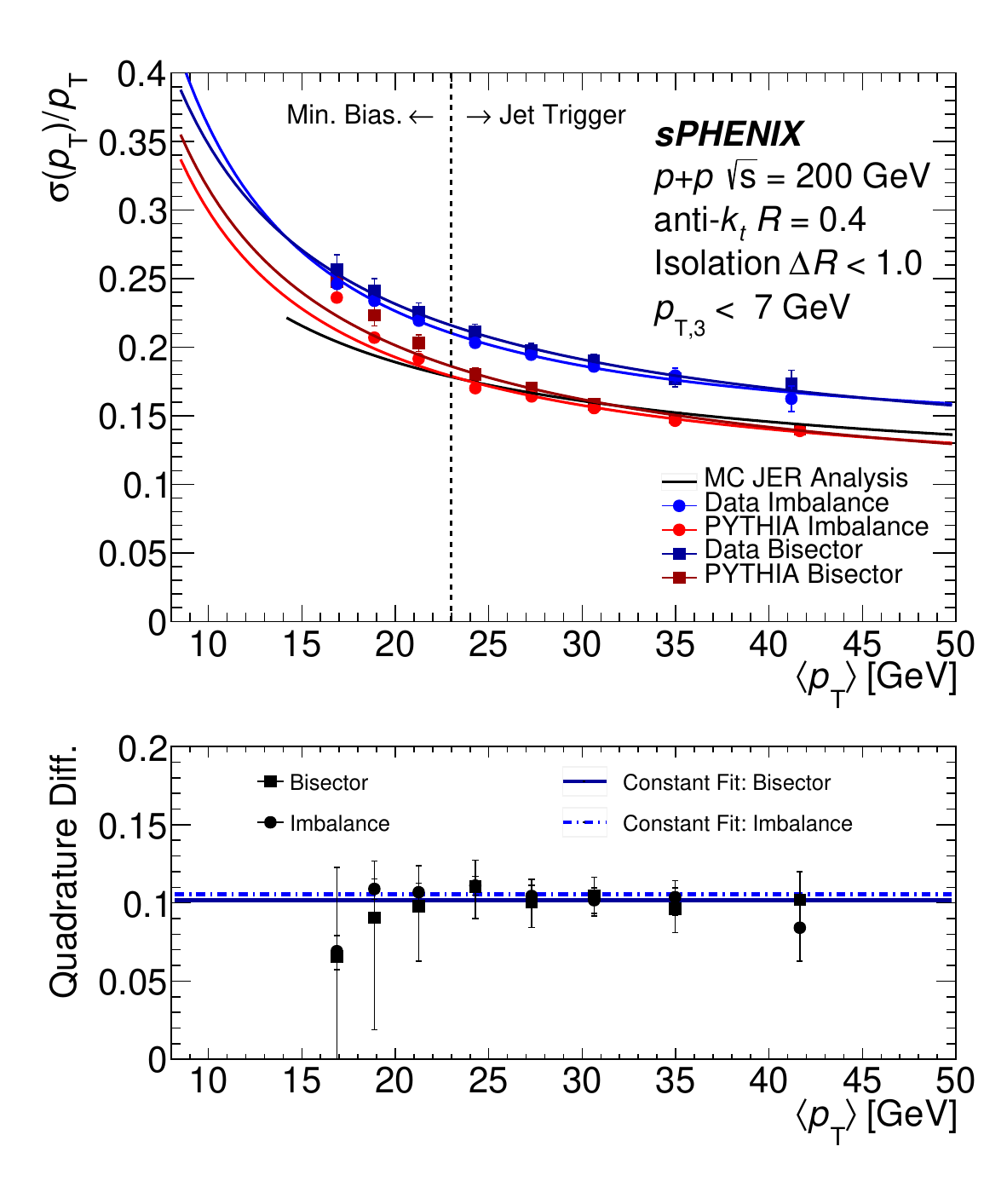}
    \caption{Top: relative jet \pt resolution, $\sigma(\pt)/\pt$, for $R = 0.4$ jets as a function of $\meanpt$, shown in data (blue) and \pythia simulation (red) for the bisector (squares) and dijet imbalance (circles) methods. Fits to the curves of the form $C\oplus{S}/\sqrt{p_\mathrm{T}}\oplus{N}/p_\mathrm{T}$ are overlaid, as well as the JER determined in simulation as a function of jet \pt\ (black line). Bottom: quadrature difference of the relative resolution between data and simulation, quantifying the additional smearing applied in the measurement.}
    \label{fig:jer_combined}
\end{figure}

Figure~\ref{fig:jer_combined} shows the extracted relative jet \pt resolution, $\sigma(\pt)/\pt$, as a function of $\meanpt$ for $R = 0.4$ jets in data and simulation, as determined by the bisector and dijet imbalance methods. The two methods are in good agreement with each other in both data and simulation, both concluding that the resolution in data is consistently larger than that in simulation. The lower panel shows the quadrature difference of the relative resolution between data and simulation, which quantifies an additional, independent contribution to the JER present in data but not described in the detector simulation. This additional component varies from approximately $10$--$12$\% depending on jet \pt.


This analysis is repeated for each jet radius parameter from $R = 0.3$ to $R = 0.8$, with the nominal additional smearing and systematic variation for each summarized in Table~\ref{tab:jer_smearing}. The nominal smearing is taken as the average of the two methods and is similar for all radii considered. The variation encompasses the difference between the methods, the statistical precision of the data, and the sensitivity to analysis choices such as the third-jet veto threshold and the split-jet exclusion region.

\begin{table}[!t]
    \centering
    \begin{tabular}{|c| c| c|}
        \hline\hline
        Jet radius ($R$) & Nominal smearing & $\pm1\sigma$ Variation \\
        \hline
        0.3 & 10.7\% & $\pm$2.4\% \\
        0.4 & 10.0\% & $\pm$1.8\% \\
        0.5 & ~9.7\% & $\pm$2.1\% \\
        0.6 & ~8.7\% & $\pm$1.5\% \\
        0.7 & ~8.7\% & $\pm$1.2\% \\
        0.8 & ~9.1\% & $\pm$2.8\% \\
        \hline\hline
    \end{tabular}
    \caption{Nominal additional JER smearing applied to the simulation and its systematic variation, for each jet radius parameter.}
    \label{tab:jer_smearing}
\end{table}

The additional resolution is incorporated into the simulation as follows. For each generated jet, a random number drawn from a Gaussian distribution with zero mean and width equal to the additional resolution term, times the particle-jet \pt is added to the \pt of the associated reconstructed jet. This smeared simulation is then used to populate the response matrix for the two-dimensional Bayesian unfolding of the $(\ptn{1}, \ptn{2})$ distributions. For the systematic uncertainty evaluation, the procedure is repeated with the smearing varied by $\pm1$ standard deviation as given in Table~\ref{tab:jer_smearing}, and the resulting changes in the unfolded \xj and corrected \dphi distributions are taken as the JER systematic uncertainty. This uncertainty is the dominant systematic for the \xj measurement across all \pt selections and cone sizes.

\subsection*{Unfolding \xj{}}

\begin{figure*}[!t]
    \centering
    \includegraphics[width=0.48\linewidth]{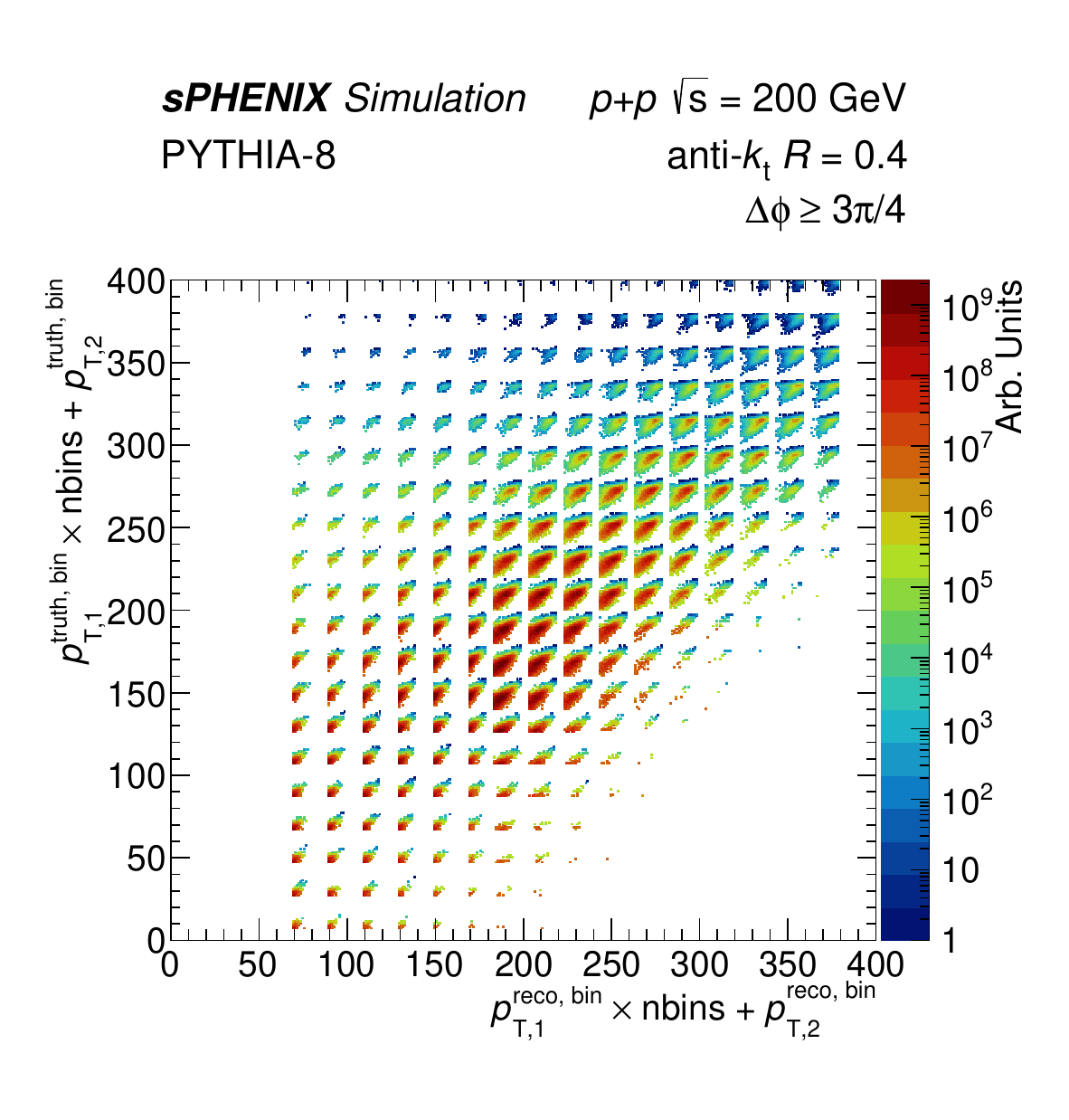}
    \includegraphics[width=0.5\linewidth]{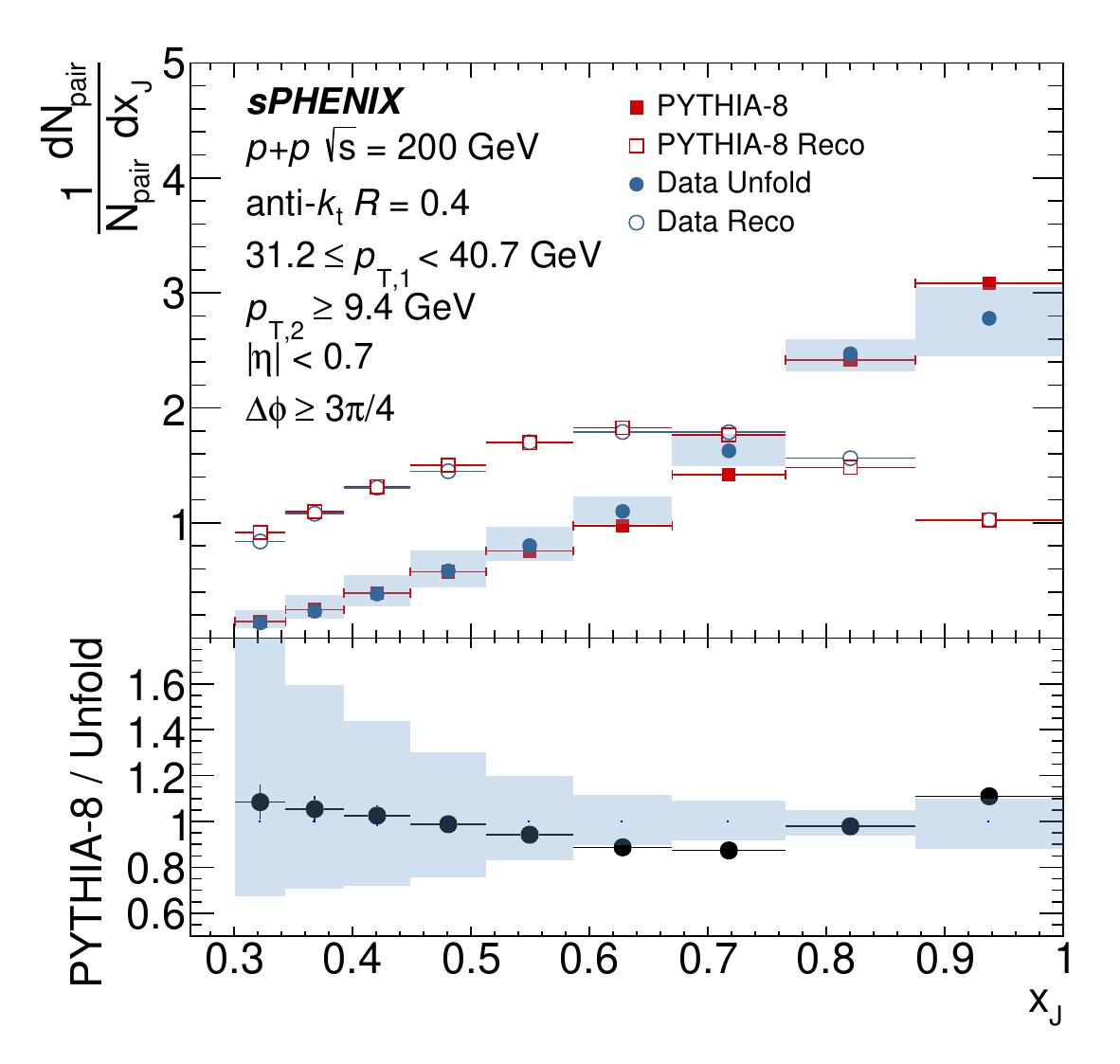}
    \caption{
    Left: Response matrix for $R = 0.4$ jets showing the association between truth- and reconstructed-level dijet $(\ptn{1}, \ptn{2})$ values in \pythia simulation. Right: 
    The top panel shows example \xj distributions from data at the reconstructed level before unfolding (open blue circles) and after unfolding (filled blue circles) in data, and at the particle (filled red squares) and reconstructed (open red squares) levels in \pythia, for $R = 0.4$ jets in the $31.2 \leq \ptn{1} < 40.7$~GeV selection. The bottom panel shows the ratio of the particle-level \pythia{} to the unfolded data. In both panels, the shaded bands denote the total systematic uncertainty on the unfolded data measurement.}
    \label{fig:method}
\end{figure*}

Figure~\ref{fig:method} shows an example of the response matrix and the \xj distribution, projected from the two-dimensional ($\ptn{1}, \ptn{2}$) distributions, before and after unfolding for an example selection.    The total systematic uncertainties on the data points are dominated by the JER uncertainty.     The lower panel shows the ratio of the \xj{} distribution in \pythia{} to data, and they are in agreement within systematic uncertainties.




\newpage

\section*{The sPHENIX Collaboration}

\begin{flushleft}
\small

M.~I.~Abdulhamid$^{13}$,
U.~Acharya\,\orcidlink{0000-0001-8560-963X}\,$^{13}$,
E.~R.~Adams$^{7}$,
G.~Adawi$^{13}$,
I.~Ahmed\,\orcidlink{0000-0003-4642-5023}\,$^{46}$,
C.~A.~Aidala\,\orcidlink{0000-0001-9540-4988}\,$^{26}$,
Y.~Akiba$^{39}$,
M.~Alfred$^{14}$,
S.~Ali$^{13}$,
A.~Alsayegh$^{10}$
% ...

\section*{Collaboration Institutes}

$^{1}$ Augustana University, Sioux Falls, South Dakota\\
$^{2}$ Baruch College, City University of New York, New York, New York\\
$^{3}$ Brookhaven National Laboratory, Upton, New York\\
$^{4}$ University of California, Los Angeles, California\\
$^{5}$ Central China Normal University, Wuhan, Hubei\\
$^{6}$ Universit\'{e} Paris-Saclay --- CEA --- IRFU, Gif-sur-Yvette, France\\
$^{7}$ University of Colorado, Boulder, Colorado\\
$^{8}$ Columbia University, New York, New York\\
$^{9}$ Debrecen University, Debrecen, Hungary\\
$^{10}$ Florida Agricultural and Mechanical University, Tallahassee, Florida\\
$^{11}$ Florida State University, Tallahassee, Florida\\
$^{12}$ Fudan University, Shanghai\\
$^{13}$ Georgia State University, Atlanta, Georgia\\
$^{14}$ Howard University, Washington, District of Columbia\\
$^{15}$ University of Illinois at Urbana-Champaign, Urbana, Illinois\\
$^{16}$ Institute for Information Industry, Taipei\\
$^{17}$ Iowa State University, Ames, Iowa\\
$^{18}$ Japan Atomic Energy Agency, Naka, Ibaraki, Japan\\
$^{19}$ Korea University, Seoul, Korea\\
$^{20}$ Lawrence Berkeley National Laboratory, Berkeley, California\\
$^{21}$ Lawrence Livermore National Laboratory, Livermore, California\\
$^{22}$ Lehigh University, Bethlehem, Pennsylvania\\
$^{23}$ Los Alamos National Laboratory, Los Alamos, New Mexico\\
$^{24}$ Lund University, Lund, Sweden\\
$^{25}$ Massachusetts Institute of Technology, Cambridge, Massachusetts\\
$^{26}$ University of Michigan, Ann Arbor, Michigan\\
$^{27}$ Morgan State University, Baltimore, Maryland\\
$^{28}$ Muhlenberg College, Allentown, Pennsylvania\\
$^{29}$ Nara Women's University, Nara, Nara, Japan\\
$^{30}$ National Central University, Taoyuan City\\
$^{31}$ National Research Nuclear University, MEPhI, Moscow Engineering Physics Institute, Moscow, Russia\\
$^{32}$ National Taiwan University, Taipei\\
$^{33}$ University of North Carolina, Greensboro, North Carolina\\
$^{34}$ Oak Ridge National Laboratory, Oak Ridge, Tennessee\\
$^{35}$ Ohio University, Athens, Ohio\\
$^{36}$ Peking University, Beijing\\
$^{37}$ Purdue University, West Lafayette, Indiana\\
$^{38}$ Pusan National University, Pusan, Korea\\
$^{39}$ RIKEN Nishina Center for Accelerator-Based Science, Wako, Saitama, Japan\\
$^{40}$ RIKEN BNL Research Center, Brookhaven National Laboratory, Upton, New York\\
$^{41}$ Rikkyo University, Toshima, Tokyo, Japan\\
$^{42}$ Rutgers University, Piscataway, New Jersey\\
$^{43}$ State University of New York, Stony Brook, New York\\
$^{44}$ Sun Yat-sen University, Guangzhou, Guangdong\\
$^{45}$ Instituto de Alta Investigaci\'{o}n, Universidad de Tarapac\'{a}, Arica, Chile\\
$^{46}$ University of Tennessee, Knoxville, Tennessee\\
$^{47}$ University of Texas, Austin, Texas\\
$^{48}$ United States Merchant Marine Academy, Kings Point, New York\\
$^{49}$ Vanderbilt University, Nashville, Tennessee\\
$^{50}$ Wayne State University, Detroit, Michigan\\
$^{51}$ Weizmann Institute of Science, Rehovot, Israel\\
$^{52}$ Yonsei University, Seoul, Korea\\

\end{flushleft}

\end{document}